\documentstyle[preprint,aps,psfig]{revtex}

\def\be{\begin{equation}}
\def\ee{\end{equation}}
\def\bea{\begin{eqnarray}}
\def\eea{\end{eqnarray}}
\begin{document}
%\begin{frontmatter}
\title{Enhanced antiproton production in  Pb(160~AGeV)+Pb reactions: 
evidence for quark gluon matter?}

\author{M.~Bleicher${}^{a,\xi}$, 
M. Belkacem${}^{b}$, S.A. Bass${}^{c}$, S. Soff${}^{d}$, H. St\"ocker${}^{d}$
}

\address{${}^a$ Nuclear Science Division,
        Lawrence Berkeley National Laboratory,\\
        Berkeley, CA 94720, USA}

\address{${}^b$ School of Physics and Astronomy,
         University of Minnesota\\
         Minneapolis, MN 55455, USA}

\address{${}^c$ National Superconducting Cyclotron Laboratory,
         Michigan State University\\
         East Lansing, MI 48824-1321, USA}

\address{${}^d$Institut f\"ur Theoretische Physik,  
         Goethe-Universit\"at,\\
         60054 Frankfurt am Main, Germany}

\footnotetext{E-mail: bleicher@nta2.lbl.gov}
\footnotetext{${}^{\xi}$ Feodor Lynen Fellow of the Alexander v. Humboldt Foundation}

%%%%%%%%%%%%%%%%%%%%%%%%%%%%%%%%%%%%%%%%%%%%%%%%%%%%%%%%%%%%%%
% You may repeat \author \address as often as necessary      %
%%%%%%%%%%%%%%%%%%%%%%%%%%%%%%%%%%%%%%%%%%%%%%%%%%%%%%%%%%%%%%

\maketitle
\begin{abstract}
The centrality dependence of the antiproton per participant ratio is studied
in Pb(160~AGeV)+Pb reactions.
Antiproton production in collisions of heavy nuclei 
at the CERN/SPS seems considerably enhanced as
compared to conventional hadronic physics, given by 
the antiproton production rates in $pp$ and antiproton annihilation in
$\overline{p}p$ reactions.
This enhancement is consistent with the observation of strong in-medium
effects in other hadronic observables and may be an indication
of partial restoration of chiral symmetry.
\end{abstract}

%\end{frontmatter}
\newpage
One of the goals of relativistic heavy-ion research is the
exploration of the phase diagram of hot and dense nuclear 
matter \cite{stoecker,qgp_rev}.
Abundances and ratios of hadrons produced in these collisions 
have been suggested as possible signatures
for exotic states and phase transitions occurring in the course of the 
reaction \cite{3,4,6}.
E.g. strangeness and antibaryon enhancement 
due to gluon fragmentation into $q\overline q$ pairs \cite{6}.
Bulk properties like temperatures, entropies and chemical potentials
of highly excited hadronic matter have been extracted from high energy
heavy ion data assuming thermal and chemical equilibrium
\cite{qgp_rev,3,4,6,braun-munzinger,cleymans,rafelski}.
However, despite the recent announcement of circumstantial evidence 
for the formation of a quark gluon plasma (QGP)
at the CERN Super-Proton-Synchrotron (SPS) \cite{cern2000}, 
unambiguous signals of a phase transition 
into an equilibrated QGP state are still missing.

The variation of hadron ratios involving anti-baryons  
as a function of centrality has been proposed in \cite{bleicherrapcom} as a 
method to distinguish equilibrium from non-equilibrium scenarios.
In models based on the assumption of
thermal equilibrium (e.g. in thermal models or hydrodynamical
models), the particle ratios are only sensitive to the 
temperature and chemical potentials at the freeze-out stage of 
the reaction. If the freeze-out criterion is universal, the ratios are
are completely insensitive to any dynamical quantities, e.g. the 
centrality of the reaction (in the extent one assumes the same baryon densities
and temperatures at freeze-out).  
Microscopic transport
theory, however, is not constrained by equilibrium assumptions.
It is used to describe the full reaction dynamics, from the 
early non-equilibrium reaction stages up to hadronic freeze-out.
On the other hand, the centrality dependence of hadron ratios provides a
sensitive tool to probe the extent at which these ratios are described by a
purely hadronic picture or by the invocation of more exotic 
pictures such as the
formation of a QGP phase or partial chiral symmetry restoration.
Thus, the centrality dependence of hadron ratios may provide a sensitive tool 
to probe the extent of the creation of a chemically equilibrated phase 
in collisions of heavy nuclei and distinguish different reaction scenarios.
The recently published data by the NA49 collaboration on the ratios 
antiprotons per participant 
as a function of the number of participants \cite{na49data} indicate a
constant enhancement of anti-proton production as compared to $pp$ results over
the whole range of centralities considered. 
The differences in the predicted centrality 
dependence among the discussed models as compared to data can be investigated
to identify whether we observe
an enhanced production of antimatter or a suppressed annihilation of
antibaryons at CERN/SPS.

For our investigation we employ the Ultra-relativistic Quantum Molecular 
Dynamics model (UrQMD) \cite{urqmd}. UrQMD is a microscopic transport
model with hadronic and constituent (di)quark degrees of freedom. 
Baryon-baryon, meson-baryon and meson-meson collisions 
lead to the formation and decay of resonances and color flux tubes. The
produced particles, as well as the incoming particles, 
rescatter in the further evolution of the system.

\begin{figure}[t]
\vskip 0mm
\vspace{-1.0cm}
\centerline{\psfig{figure=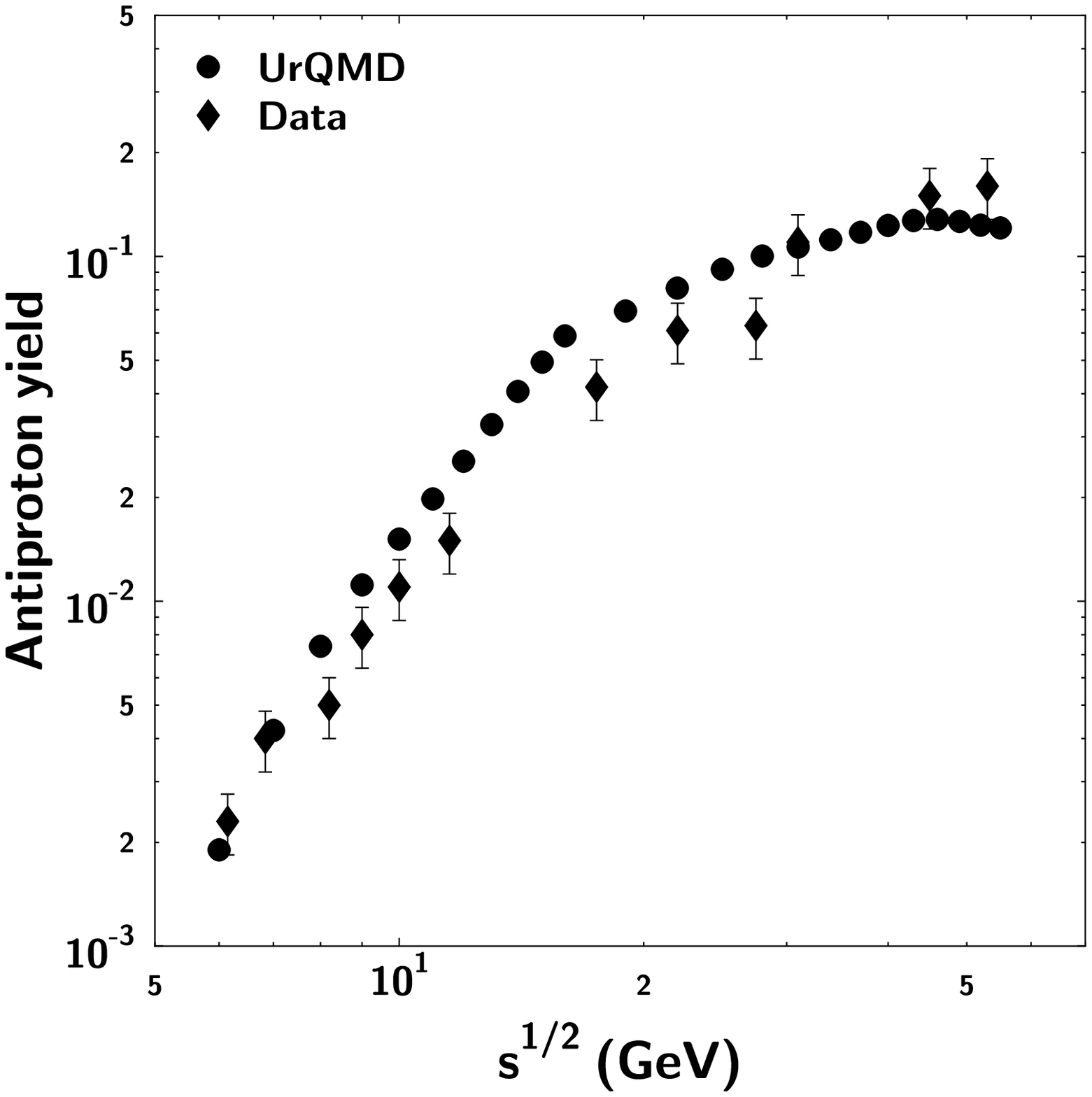,width=15cm}}
\vskip 2mm
%\vspace{-1.0cm}
\caption{Production cross sections of antiprotons in pp as a function of
energy. 
\label{prod}}
\end{figure}

Let us start by investigating the general features of antiproton production
and annihilation in UrQMD:
Antiprotons are produced via the decay of color flux tubes and in antiresonance
decays (e.g. $\overline{\Delta^+} \rightarrow \overline p + \pi$). 
Since the collision dynamics in nucleus-nucleus is a
convolution from high and  low energy elementary collisions, 
it is important to verify that the energy dependence of the antiproton 
production is reasonable.
Fig. \ref{prod} shows the multiplicity of antiprotons in inelastic $pp$
interactions as a function of center of mass energy ($\sqrt s$) compared
to data.
The UrQMD predictions, depicted as full circles, overestimate 
the experimental data (shown as full diamonds) by 50\% in the SPS domain
($\sqrt s =20$~GeV).
In line with data, the string model shows a strong 
increase of antiproton production
with collision energies starting from a threshold of $4m_{\rm proton}$. 
At higher energies,
the production cross sections levels off at $\overline p$ 
multiplicities of 0.1-0.2. However, the antiproton production will 
increase further at even higher energies, when the double antibaryon-baryon 
channels become populated. 

\begin{figure}[t]
\vskip 0mm
\vspace{-1.0cm}
\centerline{\psfig{figure=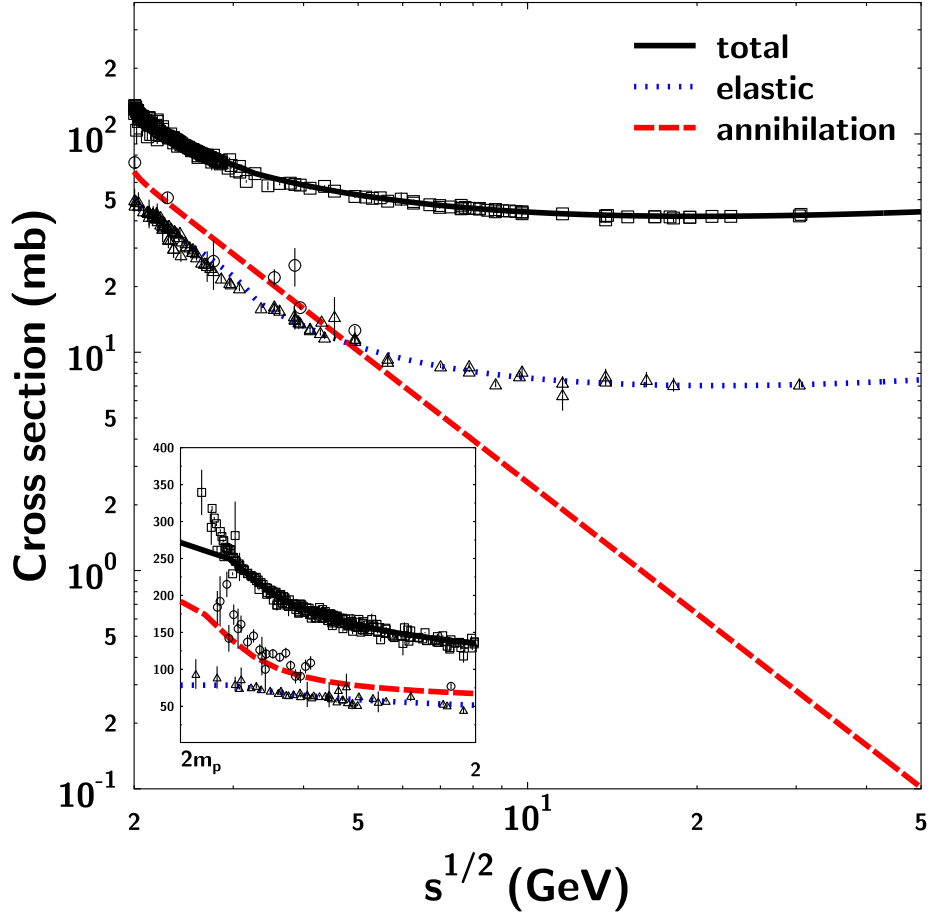,width=15cm}}
\vskip 2mm
%\vspace{-1.0cm}
\caption{Elementary antiproton-proton total, elastic and 
annihilation cross section as a function energy compared to data (symbols).  
Note that the sum of the 
annihilation cross section and the inelastic cross section is smaller than the
total cross section. The difference $\Delta \sigma = \sigma_{\rm total} -
\sigma_{\rm elastic} - \sigma_{\rm annih.}$ is assumed to be a pure inelastic
cross section in $\overline p p$ in analogy to $pp$.\label{annih}}
\end{figure}

In massive nuclear collisions, however, not only the production of antiprotons
must be treated, but also antibaryon absorption can be important. In UrQMD the
antiproton annihilation is modeled via the annihilation of quark-antiquark pair
and the formation and subsequent
decay of two color flux tubes with baryon number zero (for 
details see \cite{urqmd}).
Fig. \ref{annih} confronts the UrQMD implementation of the
antiproton-proton cross sections as a function of center of mass energy with
experimental data \cite{pdg}. 
The full line shows the total $\overline p p$ cross section
(data is shown as squares), the dotted line shows the 
elastic cross section (data
as triangles) and the dashed line depicts the annihilation cross 
section in UrQMD. 
In addition, the small inlay gives a detailed view of the cross sections from
$2m_{\rm proton}$ to $\sqrt s =2$~GeV. This region is of special interest,
since the antiproton absorption strongly increases towards low center of mass
energies. Overall, a reasonable description of the antiproton-proton 
interactions over the energy range $\sqrt s \leq 5$~GeV (most relevant for SPS)
is obtained within the UrQMD model.  Note that the sum of the 
annihilation cross section and the inelastic cross section is smaller than the
total cross section. The difference $\Delta \sigma = \sigma_{\rm total} -
\sigma_{\rm elastic} - \sigma_{\rm annih.}$ is assumed to be a pure inelastic
cross section in $\overline p p$ in analogy to $pp$.

\begin{figure}[t]
\vskip 0mm
\vspace{-1.0cm}
\centerline{\psfig{figure=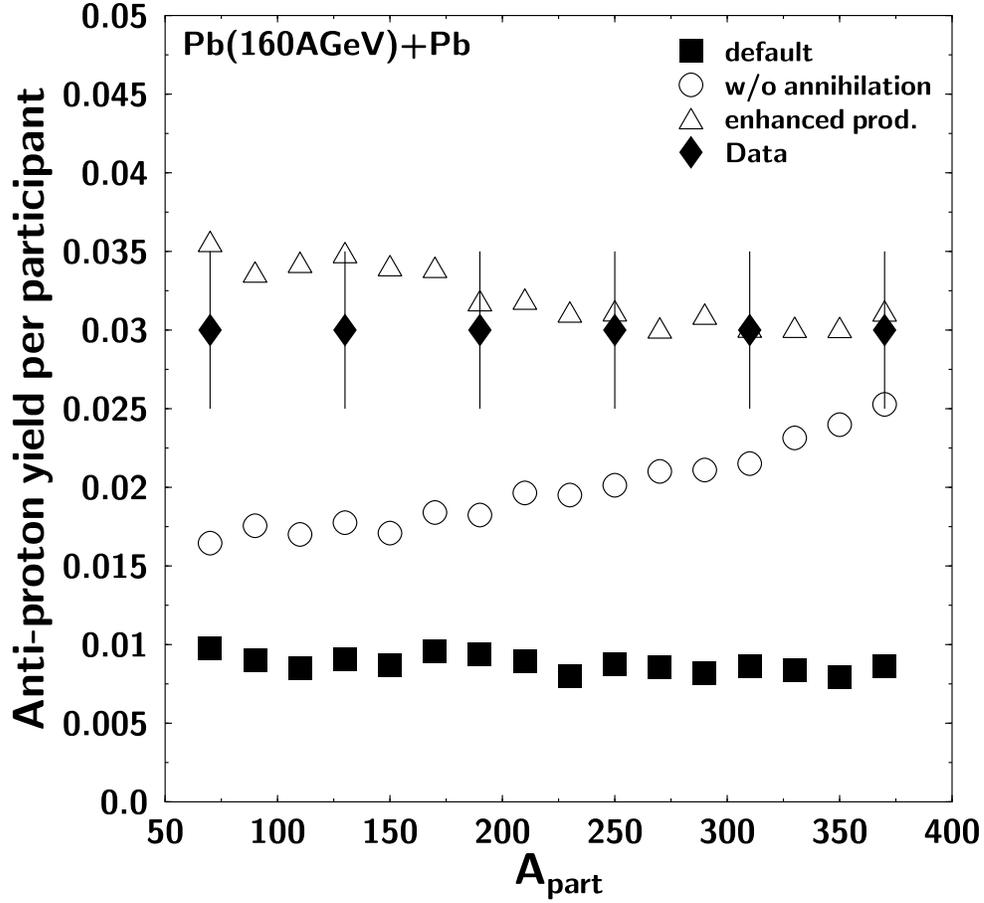,width=15cm}}
\vskip 2mm
%\vspace{-1.0cm}
\caption{Scaling of the antiprotons production with the number of participating
nucleons in Pb(160 AGeV)+Pb collisions. Full diamonds depict 
experimental data, full squares show the standard UrQMD calculation,
open circles show the UrQMD calculation with rescattering switched off, while
the open triangles show UrQMD simulations with enhanced antibaryon production
and standard antibaryon absorption. 
\label{scaling}}
\end{figure}

Let us now turn towards the dynamics of nucleus-nucleus collisions 
at 160~AGeV. In addition to the $pp$ case, meson-meson and 
meson-baryon interactions may also
lead to the excitation of color flux tubes and their subsequent decay into
baryon-antibaryon pairs. E.g. the channel $\rho\rho \rightarrow \overline B B$
constitutes a new and unexplored production channel in AA collisions.
On the other hand, 
the high baryon densities reached at 
central rapidities in nucleus-nucleus interactions can lead to an 
increased absorption of antiprotons.
Fig. \ref{scaling} shows the $\overline{p}/$participant\footnote{%
The number of participating nucleons in this paper is defined as:\\
$A_{\rm part}= A_1+A_2 - \Sigma \, (\mbox{Nucleons with } p_T\leq 270$~MeV).
This prescription yields a reasonable parametrization of the experimental data
on $A_{\rm part}$ as can be seen by comparison to Ref. \cite{cooper}.
}%
ratio for different centralities (given by the number of participating 
nucleons $A_{\rm part}$).
Full squares show the standard UrQMD prediction of 
the $\overline{p}/A_{\rm part}$ ratio, while the data are depicted as full 
diamonds. A constant ratio over all centralities is observed both in
experiment and calculation. In view of the large changes in the reaction
dynamics when going from peripheral to central collisions the lack 
of $A_{\rm part}$ dependence in the $\overline{p}/A_{\rm part}$ 
ratio seems surprising. However, the UrQMD model calculations 
{\it under}estimate the data by a factor of 3! Considering the 50\%
overestimate in the elementary production channel (Fig. \ref{prod}) this
drastic deviation can only be explained by
\begin{enumerate}
\item strongly enhanced production,
\item strongly suppressed annihilation.
\end{enumerate}

Within the UrQMD model, we are able to further investigate the cause
of this behavior:
since the elementary production and absorption cross sections are in line with
the data, we expect one of the following scenarios to take place in AA 
collisions:
\begin{itemize}
\item A suppressed antiproton annihilation in dense matter, e.g. due to pion
clouds which prevent annihilation. Such a screening effect has
been speculated upon by \cite{kahana}.
\item
An enhanced production of antiprotons in AA, compared to $pp$ extrapolations.
This has been predicted by \cite{6,uheinz} as a signature of a QGP phase
transition. It is interesting to note that studies by Koch et al. \cite{6} and 
Ellis et al., \cite{ellis} predict a factor $2.5-10$ enhancement of the
antibaryon production due to QGP formation.
This $\overline{p}$ source is also in line with recent 
measurements on the (anti-)hyperon enhancement 
in AA \cite{strange_data,ssoff}. 
A remark here is in order: As stated above, the flat enhancement of 
$\overline{p}/A_{\rm part}$ as compared to $pp$ results at all centralities
is amazing. If the observed enhancement in the data is explained by the 
transition to the QGP, the same data indicate the formation
of the QGP even in peripheral collisions (or in the 
smaller Sulphur-Sulphur system) where the formation of  
such a state seems less favourable due to the lower energy- and 
baryon-densities reached in these collisions.
\end{itemize}
To study the influence of these effects -- and possibly distinguish between
them --  we alternatively incorporate the above mentioned effects into UrQMD
and compare the results to the default scenario and the data.

First, Fig. \ref{scaling} (open circles) shows the calculation with 
antibaryon annihilation turned off, as an extreme case of the 
screening assumption. This option leads to a reasonable
description of the most central Pb+Pb interactions. However, the scaling of the
$\overline{p}/A_{\rm part}$ ratio exhibits an
increase of the ratio towards central collisions not in line with the NA49 
data. Clearly, within the hadronic/string physics of UrQMD,
antiproton production is enhanced in central collisions - annihilation
is required to compensate this enhancement and obtain the flat
$\overline p /A_{\rm part}$ ratio vs. centrality. We note also that the
annihilation of anti-protons plays a counter-balancing role with anti-baryon 
production in secondary scattering (meson-baryon and meson-meson) to maintain 
the $\overline{p}/A_{\rm part}$ ratio constant versus $A_{\rm part}$.

In order to test the deconfinement hypothesis, we perform a calculation
with full antiproton absorption, but with an enhanced $\overline p$ 
production cross section. Normally, one might think to enhance the 
cross-section smoothly from the $pp$ cross-section at peripheral reactions 
to some large value for central collisions. In the present study, we chosed 
a constant enhancement, adjusted to the most central collisions.
This scenario, shown as open triangles in Fig. \ref{scaling}, 
leads to a flat $\overline{p}/A_{\rm part}$ ratio as function of 
centrality, in line with the data.
This increase in antiproton production is due to an increase of the
string tension by a factor of 2.6. 
This increase of the string tension results in an enhanced
di-quark -- anti-diquark production probability due to the Schwinger
formula:
\begin{equation}
\gamma_{qq}=\frac{P(qq\bar{q}\bar{q})}{P(q\bar{q})}=\exp
\left(- \frac{\pi
(m_{qq}^2-m_q^2)}{\kappa}\right)\,.
\label{gamm}
\end{equation}
leading to an increase of the diquark 'suppression' parameter \footnote{%
To be very specific: The light quark mass used in UrQMD is $m_q=0.223$~GeV and
the diquark mass $m_{qq}=2m_q$, the string tension is $\kappa 
= 1 \mbox{ GeV/fm} = 0.2 \mbox{ GeV}^2$, resulting in $\gamma_{qq} = 0.095$. 
These values provide a reasonable description of the $\overline p$ 
production in $pp$ as demonstrated above. However, to describe 
the AA data, $\gamma_{qq}$ needs to be enhanced to 0.4. Using Eq. \ref{gamm}
this leads to an effective $\kappa^{'}$ of $\kappa^{'} = - 3\pi m_q^2 / 
\left({\rm ln} \gamma_{qq}\right) = 0.511 \mbox{ GeV}^2$. Thus, an increase 
in the string tension by a factor 2.6.}   
$\gamma_{qq}$ from
0.1 to 0.4. This spectecular increase of the string tension has recently been
employed to reproduce the observed (anti-)hyperon enhancement at SPS 
energies \cite{ssoff}. It is consistent with the assumption of 
the onset of partial restoration of chiral symmetry which might lead to a
decrease of the constituent quark masses towards current quark masses. 
The increased string tension can be alternatively motivated by the assumption
of overlapping color flux tubes (''ropes''). Such superposition 
of the color electric fields
can yield enhanced particle production \cite{biro84,sor92}.
In particular, heavy quark flavors and diquarks are dramatically
enhanced  \cite{sor92,gyulassy90,gerland95}. 

While there seems to be a strong enhancement
of the antiproton production established, which causes the observed centrality
dependence of the $\overline p /A_{\rm part}$ ratio, the exact cause
of the enhancement remains ambiguous:
several different mechanisms can
lead to enhanced antiproton production: overlapping color flux tubes,
QGP formation, reduced hadron masses
due to partial restoration of chiral symmetry 
and multi-particle interactions at high densities,
e.g. $\pi\pi\pi\pi\pi\rightarrow \overline B B$. 
The absence of the latter type of processes in the UrQMD model (and in all
other hadronic and string models, e.g. HIJING, RQMD, VENUS, etc.)
leads to a violation of detailed balance. It yields   
significant deviations from the expected properties of an ideal 
hadron gas in the  equilibrium limit of UrQMD 
(infinite volume and infinite time at fixed energy density) \cite{belkacem}. 
For the fast-changing environment and reaction
dynamics of a relativistic heavy-ion collision, however, it remains 
open whether the inclusion of these phase-space suppressed processes would
change the results significantly. On the other hand, all the 
above mentioned mechanisms
have a strong centrality dependence and while they might be justified and used
to explain the enhancement observed in the data
for central collisions, their influence at peripheral impact parameters is
questionable.

\begin{figure}[t]
\vskip 0mm
\vspace{-1.0cm}
\centerline{\psfig{figure=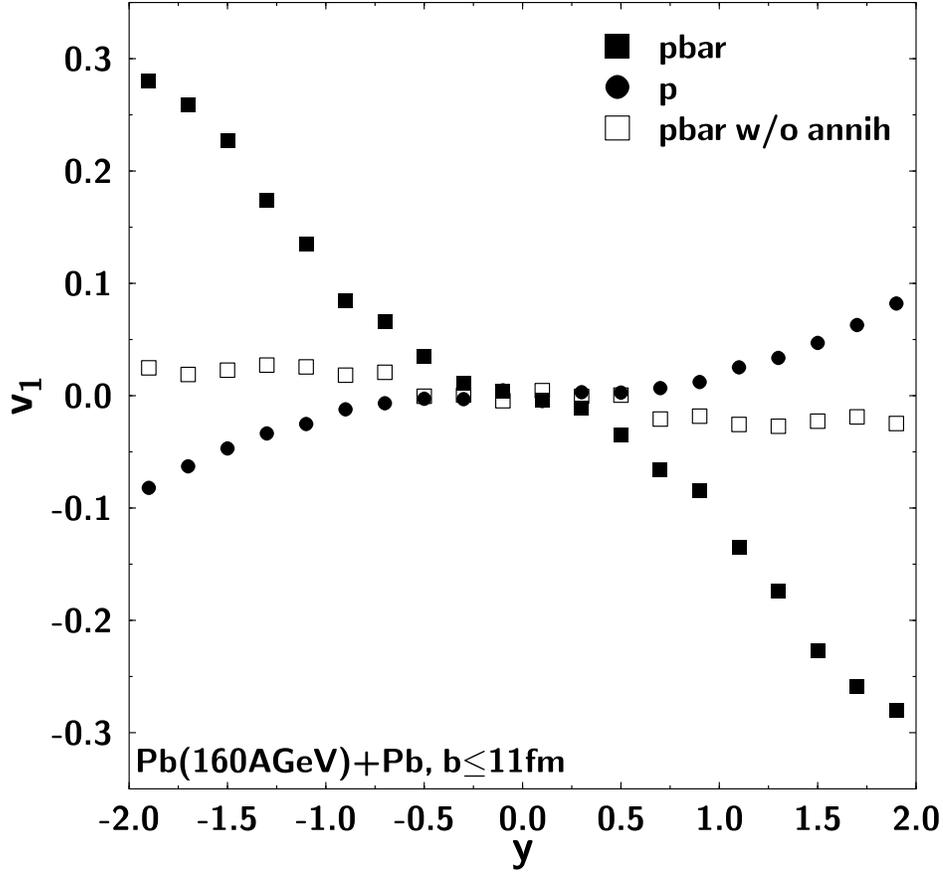,width=15cm}}
\vskip 2mm
%\vspace{-1.0cm}
\caption{Anisotropic flow parameter $v_1$ in Pb(160 AGeV)+Pb, $b\leq 11$~fm 
as a function of rapidity for protons (full circles), antiprotons (full
squares) and antiprotons with annihilation switched off (open squares). 
\label{v1y}}
\end{figure}

Further understanding of the antiproton dynamics in dense matter 
can be obtained by studying the anisotropic flow parameter $v_1$ as shown 
in Fig. \ref{v1y} for Pb(160 AGeV)+Pb interaction at impact 
parameters $b\leq 11$~fm.  
This provides an independent and direct check of the in-medium absorption cross
section. 
With antiproton absorption (full squares) a strong anti-flow \cite{jahns} 
of antiprotons is predicted. The strength of the 
flow is 2-3 times stronger than 
for the proton flow (shown as full circles). If annihilation is suppressed, 
the anti-flow of antiprotons nearly vanishes (open squares).
Thus, an experimental study of the anti-flow of antiprotons can provide 
direct access to the $\overline p p$ annihilation cross section in 
dense matter.

The production and absorption of antiprotons in elementary collisions has been
compared to data and reasonable agreement has been found. 
It has been shown that within the UrQMD model these cross sections 
result in an underprediction of the antiproton yield
in Pb+Pb at 160 AGeV at all centralities.
Suppressing of antiproton annihilation in AA collisions 
does not seem to be a possible cause for the observed centrality 
dependence of the $\overline p /A_{\rm part}$ ratio. It is
demonstrated that the measured data are consistent with an enhanced
production cross section of $\overline p$'s in nucleus-nucleus collisions.
The study of the anti-flow of antiprotons 
can give a definitive answer on the antibaryon absorption
cross section in hot and dense matter. 
%We also note that the constant
%enhancement shown by the data of the $\overline p /A_{\rm part}$ ratio as 
%compared to $pp$ results over all centralities considered is quite surprising.
%While this enhancement might be explained for central collisions by different 
%mechanisms such as QGP formation, overlapping color flux tubes and reduced 
%hadron masses due to partial restoration of chiral symmetry, using these
%mechanisms to explain the data for peripheral collisions is strongly
%questionable.

\section*{Acknowledgements}
The authors would like to thank L. Gerland for fruitful discussions. 
M. Bleicher is supported by the A. v. Humboldt Foundation. 
S.A.B. acknowledges financial support from the U.S. National Science
Foundation, grant PHY-9605207. M.Belkacem was supported by the U.S. Department
of Energy under grant No. DE-FG02-87ER40328.
This work is supported by the BMBF, GSI, DFG and the Graduiertenkolleg
'Theoretische und experimentelle Schwerionenphysik'.
This research used resources of the
National Energy Research Scientific Computing Center (NERSC).

\newpage

\newpage

\newpage

\newpage

\end{document}